\documentstyle[11pt,newpasp,twoside,epsf]{article}

\def\edcomment#1{\iffalse\marginpar{\raggedright\sl#1\/}\else\relax\fi}
\reversemarginpar

\begin{document}

\title{Analysis of On-Orbit ACIS Squeegee Mode Data}
\author{Shanil N. Virani and Paul P. Plucinsky}
\affil{Harvard-Smithsonian Center for Astrophysics, 60 Garden St.,
MS-70, Cambridge, MA 02138}


\author{Catherine E. Grant and Beverly LaMarr}
\affil{Center for Space Research, Massachusetts Institute of Technology, 70 Vassar St.,
Cambridge, MA 02139}


\begin{abstract}
The MIT and CXC ACIS teams have explored a
number of measures to ameliorate the effects of radiation damage
suffered by the ACIS FI CCDs. One of these measures is
a novel CCD read-out method called ``squeegee mode''. A variety of
different implementations of the squeegee mode have now been tested on
the I0 CCD. 

Our results for the fitted FWHM at Al-K$\alpha$ and Mn-K$\alpha$ 
clearly demonstrate  that
all the squeegee modes provide improved performance in terms of
reducing CTI and improving spectral resolution. Our analysis of the 
detection efficiency shows that the so-called squeegee modes
``Vanilla'' and ``Maximum Observing Efficiency'' provide the same 
detection efficiency as the standard clocking, once the decay in the 
intensity of the radioactive source has been taken into account.  The
squeegee modes which utilize the slow parallel transfer (``Maximum 
Spectral Resolution'', ``Maximum Angular Resolution'', and ``Sub-Array'')
show a significantly lower detection efficiency than the 
standard clocking. The slow parallel transfer squeegee modes exhibit 
severe grade migration from flight grade 0 to flight grade 64 and a 
smaller migration into ASCA g7.  The latter effect can explain some 
of the drop in detection efficiency.  

There are a few observational penalities to consider in using a squeegee
mode. Utilizing any squeegee mode causes a loss of FOV near the
aimpoint (4 to 16$''$ strips along the full length of the CCDs), as well
as the attendant dead-time increase. Secondly, the cost of the
software implementation and its testing will be significant. Lastly,
each squeegee mode ``flavor'' would require lengthy,
mode-specific calibration observations. Therefore, since an efficacious, 
ground-based CTI corrector algorithm is now available (see paper by
Plucinsky, Townsley, \textit{et al.} in this proceedings), a scientific judgment 
will have to be made to determine which, if any, squeegee modes should
be developed and calibrated for use by Chandra observers.

\end{abstract}

\section{Introduction}

   The ACIS Operations team, combining elements from the CXC and the
ACIS MIT$/$IPI team, has been developing and testing the new squeegee modes
since the Spring of 2000. In squeegee mode, charge is collected in the 
top few rows of the CCD and then swept across the imaging array once per readout,
thus filling some of the radiation-induced electron traps
that cause degraded performance.
The development of a novel method of reading
out the ACIS CCDs was first developed and tested on ground CCDs that
are similar to flight CCDs
(see Prigozhin, \textit{et al.} 2000 for a characterization
of the radiation damage of ACIS CCDs). This new mode ameliorates some
of the effects caused by the radiation damage suffered early
in the mission by the FI CCDs. For a more
comprehensive discussion on the design and clocking method of squeegee
mode, see Bautz and Kissel (2000).  A description of the different squeegee
modes is included in an internal MIT$/$ACIS Memo by Bautz and Grant (2000).  
That memo also presents an analysis of the squeegee modes and
describes the trade-offs in choosing between the various squeegee modes.  
In this paper, we present the results of the various squeegee measurements on
the I0 CCD.

\begin{table}[tbp]
\centering
\caption[ ]{{\bf List of I0 Squeegee Tests } }
\label{datasets}
\begin{tabular}{|l|l|l|l|l|}
\hline
Test & OBSID   & Date  & Exp(s)    & Description \\
\hline
\hline
Control   & 62895  & 20 Feb 2000 & 9,023.4
& I0, 3.2s, standard clocking\\
& & & & ($40\mu$s par xfr, no 2x2 sum)\\
& & & & ``Control'' non-Squeegee \\
\hline
L1  & 62042 & 30 May 2000 & 8,225.5 
& I0, 3.3s, 16 row sq, 32 row exc win, no 2x2 sum, \\
& & & & $40\mu$s par xfr, 24 flushes, 1010 rev clks \\
& & & & ``Vanilla Squeegee'' \\
\hline
L2 & 62019    & 21 Jun 2000  & 8,623.6 
& I0, 1.9s, 16 row sq, 24 row exc window, 2x2 sum, \\
& & & & $320\mu$s par xfr, 24 flushes, 1026+32 rev clks  \\
& & & & Maximum Spectral Resolution \\
\hline
L4 & 62007    & 01 Jul 2000 & 8,517.3 
& I0, 3.3s, 2 row sq, 8 row exc win, no 2x2 sum, \\
& & & & $320\mu$s par xfr, 24 flushes, 1026+32 rev clks  \\
& & & & Maximum Angular Resolution \\
\hline
L7 & 61981 & 30 Jul 2000 & 8,551.5 
& I0, 1.8s, 2 row sq, 8 row exc win, 2x2 sum, \\ 
& & & & $40\mu$s par xfr, 24 flushes, 
1026+32 rev clks  \\
& & & & Maximum Observing Efficiency \\
\hline
L9 & 61949  & 30 Aug 2000   & 8,356.9
& I0, 1.2s, 2 row sq, 8 row exc win, no 2x2 sum, \\ 
& & & & $320\mu$s par xfr, 24 flushes, 1026+32 rev clks, \\
& & & & 256 row sub-array  \\
\hline
\end{tabular}

\end{table}

This analysis utilized the standard level 0 and level 1 data products
produced by the CXC Data System.  The datasets included in this
analysis are listed in Table 1.  The control run (OBSID
62895) was a long charge transfer inefficiency (CTI) measurement.
There are five
``flavors'' of Squeegee discussed in this paper: L1 ``Vanilla'', L2
``Maximum Spectral Resolution'', L4 ``Maximum Angular Resolution'', L7
``Maximum Observing Efficiency'', and L9 ``Sub-Array near the
aimpoint''.  Table 1 summarizes the important parameters which distinguish
one squeegee mode from another: static integration time,
number of squeegee rows (2 or 16), 
number of rows in the exclusion window (8 or 24 or 32), on-chip
summing (yes or no), parallel transfer time ($40\mu$s or $320\mu$s),
number of frame flushes (always 24 for these squeegees), and number
of reverse clocks (1010 or 1026+32).  In addition, the table includes
the average exposure time for these measurements.  With all of these
squeegee modes, the exposure time varies from row-to-row.  The number
listed is the correct exposure time for the middle of the CCD or
for the middle of the sub-array for squeegee L9.  In the analysis
presented in this paper, the row-to-row variation in 
exposure time has been included by computing the exposure for each of 
the 32 row elements.

\section{Methodology}

	Data from each I0 measurement was separated by node and then
further reduced to thirty-one, 32-row region files. Spectra files were
generated from the CXC level 1 events files. Events were selected in
the {\tt ASCA} g02346 grade set and were filtered on node and {\tt
chipy} coordinates.  The gain was computed using the
so-called ``local-gain'' method, meaning the fitted peak in each 32
row element is used to compute a conversion from ADUs to eV for that
element. The prominent lines of the internal calibration source were
then fit using Gaussians; the figure of merit employed was the
C-statistic.

\section{Results and Analysis}

 The results of this analysis are presented in Figures~1 and 2.
Figure 1 shows the fitted values of the FWHM in eV for the I0 CCD at
1.5 keV (Al-K$\alpha$) at a focal plane temperature of -110 C, -120 C,
and at -120 C using the ``Vanilla'' squeegee mode. For comparative
purposes, the S3 CCD FWHM at -120 C is also overlaid. Figure 2 shows
the fitted values of the FWHM in eV for the I0 CCD at
5.9 keV (Mn-K$\alpha$) at a focal plane temperature of -110 C, -120 C,
and at -120 C using the ``Vanilla'' squeegee mode. For comparative
purposes, the S3 CCD FWHM at -120 C is also overlaid. What is clear
from both plots is that squeegee mode improves the spectral resolution
of I0 CCD compared to standard clocking.

Figures 3 and 4 displays the quantities {\tt ASCA} {\tt g0234/all grades}, 
{\tt ASCA} {\tt g02346/all grades}, {\tt ASCA} {\tt g7/all grades} 
versus
row number, where ``all grades'' means all events which
pass the on-board filtering and make it into telemetry, for the
Al-K$\alpha$ and Mn-K$\alpha$ lines.  All of these measurements 
were executed using an
upper event amplitude
cutoff of 3750~ADUs and a grade filter which
rejected flight grades 24, 66, 107, 214, and 255. Figure 5 is a plot
of the detection efficiency for each line as a function of row number.
We have included a statistical error bar for the count rate plots
at the location of the first data element which is representative of
the uncertainties in these measurements. In producing Figure 5, we
have corrected for the decrease in intensity of the radioactive source
as the data span 6 months. See 3.4 for further discussion.

\begin{figure}[htb]
\begin{center}
\epsfxsize=4.0in
\epsfysize=4.0in
\hspace*{0pt}
\epsffile{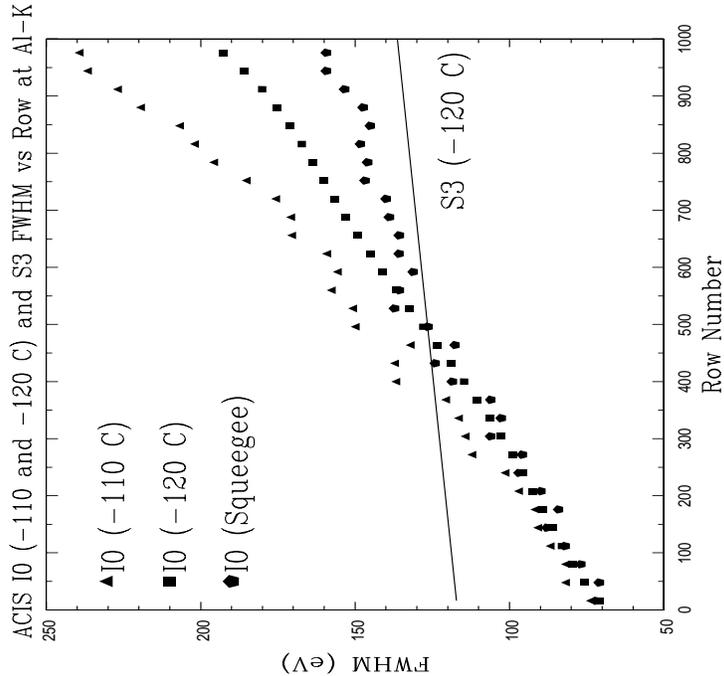}
\caption{FWHM at Al-K$\alpha$ on I0 and S3}
\end{center}
\end{figure}

\begin{figure}[htb]
\begin{center}
\epsfxsize=4.0in
\epsfysize=4.0in
\hspace*{0pt}
\epsffile{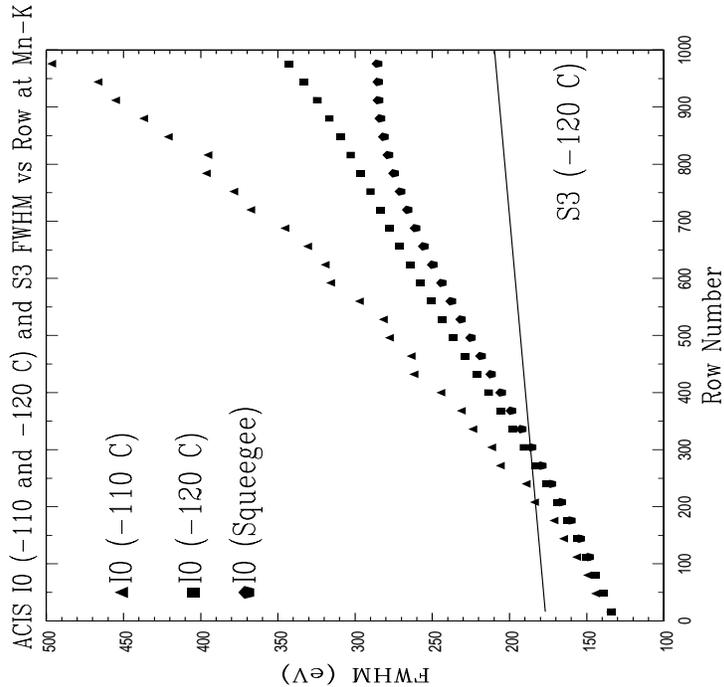}
\caption{FWHM at Mn-K$\alpha$ on I0 and S3}
\end{center}
\end{figure}

\subsection{Al results}

\begin{figure}[htb]
\begin{center}
\epsfxsize=3.0in
\epsfysize=3.0in
\hspace*{0pt}
\epsffile{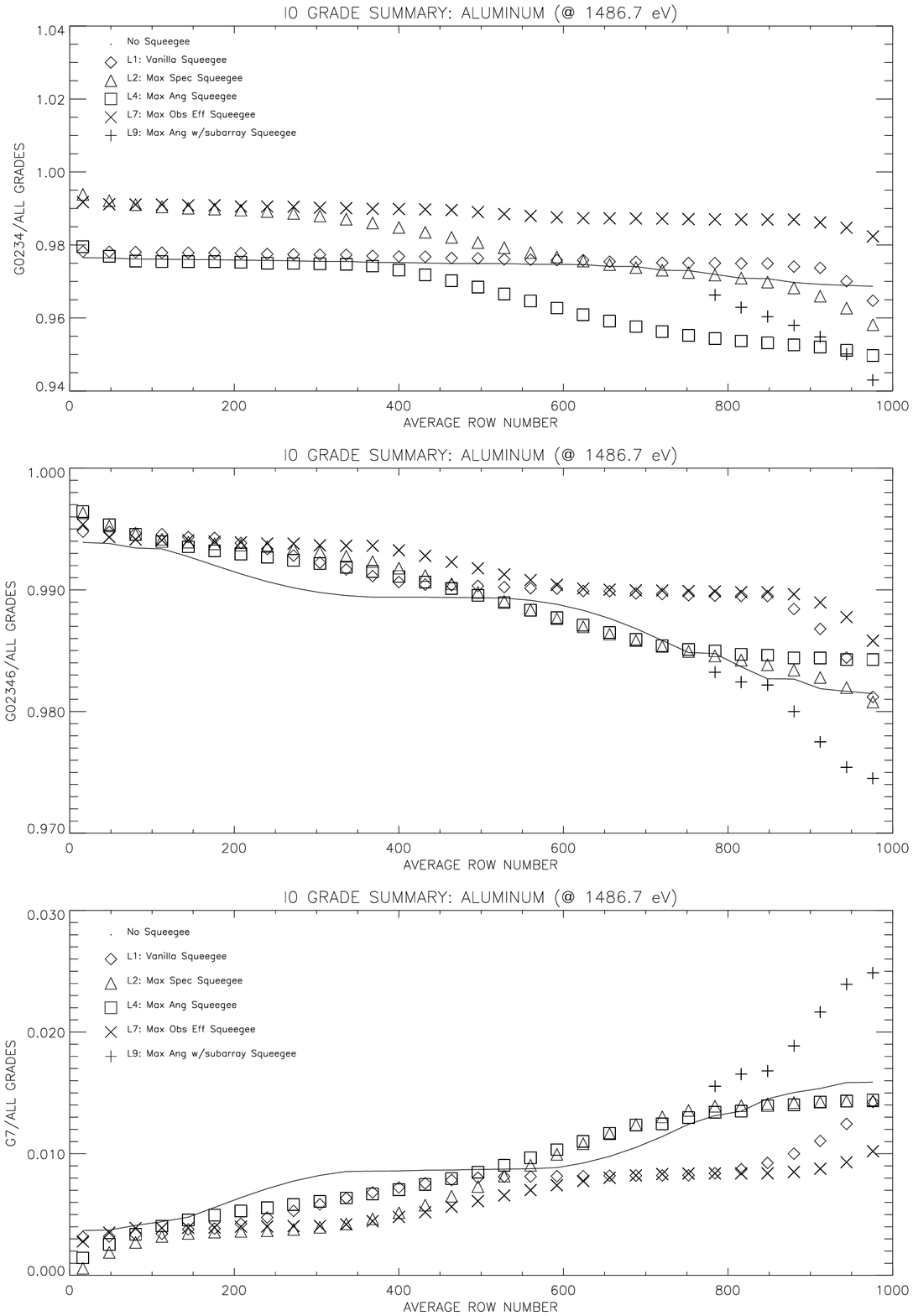}
\caption{Squeegee mode grade summary plots for the Al-K$\alpha$ line for
   all squeegee modes.}
\end{center}
\end{figure}

Figure 3 shows the grade distribution as a function of row
number for the squeegee measurements and the control run.
The enhanced 
CTI of the FI CCDs causes events to migrate from {\tt ASCA} grades g0234
to higher grades.  This can be seen in the control run and
the squeegee runs.  The effect is quite small at Al-K$\alpha$,
never more than 3\% for the grade sets listed here.
Squeegees L2 and L4 show a very
different behavior with regards to grade migration when the
individual {\tt {\tt ASCA}} grades are examined. Analysis of the g0
events as a function of row number show that the percentage of grade g0 events is
dropping from $\sim90\%$ in the first resolution element to $\sim20\%$
by the top of the CCD.  {\tt ASCA} grade 2 is the beneficiary of these counts as
it increases from its nominal value of 5\% to over 80\% of the events
at the top of the CCD. {\tt ASCA} grade 2 is composed of flight grades 2
(down splits) and 64 (up splits). A further examination of the data
reveals that all of the migrating events are going into flight grade
64.  The one characteristic which the squeegees L2 and L4 have in 
common is the slow parallel transfer.  Lastly, L2 and L7 both have much
higher g0234$/$all grades ratios than the other modes because both L2
and L7 utilize 2x2 summing.

We suggest that this severe grade migration provides a clue to the cause of the
lower detection efficiency seen in Figure 5.  A common cause of
reduced efficiency in a given grade combination is that events are
migrating to grades outside of the chosen set. We will defer a
full discussion of this until Section 3.4 in which the
decay in the intensity of the radioactive source is included.

\subsection{Mn results}

\begin{figure}[htb]
\begin{center}
\epsfxsize=3.0in
\epsfysize=3.0in
\hspace*{0pt}
\epsffile{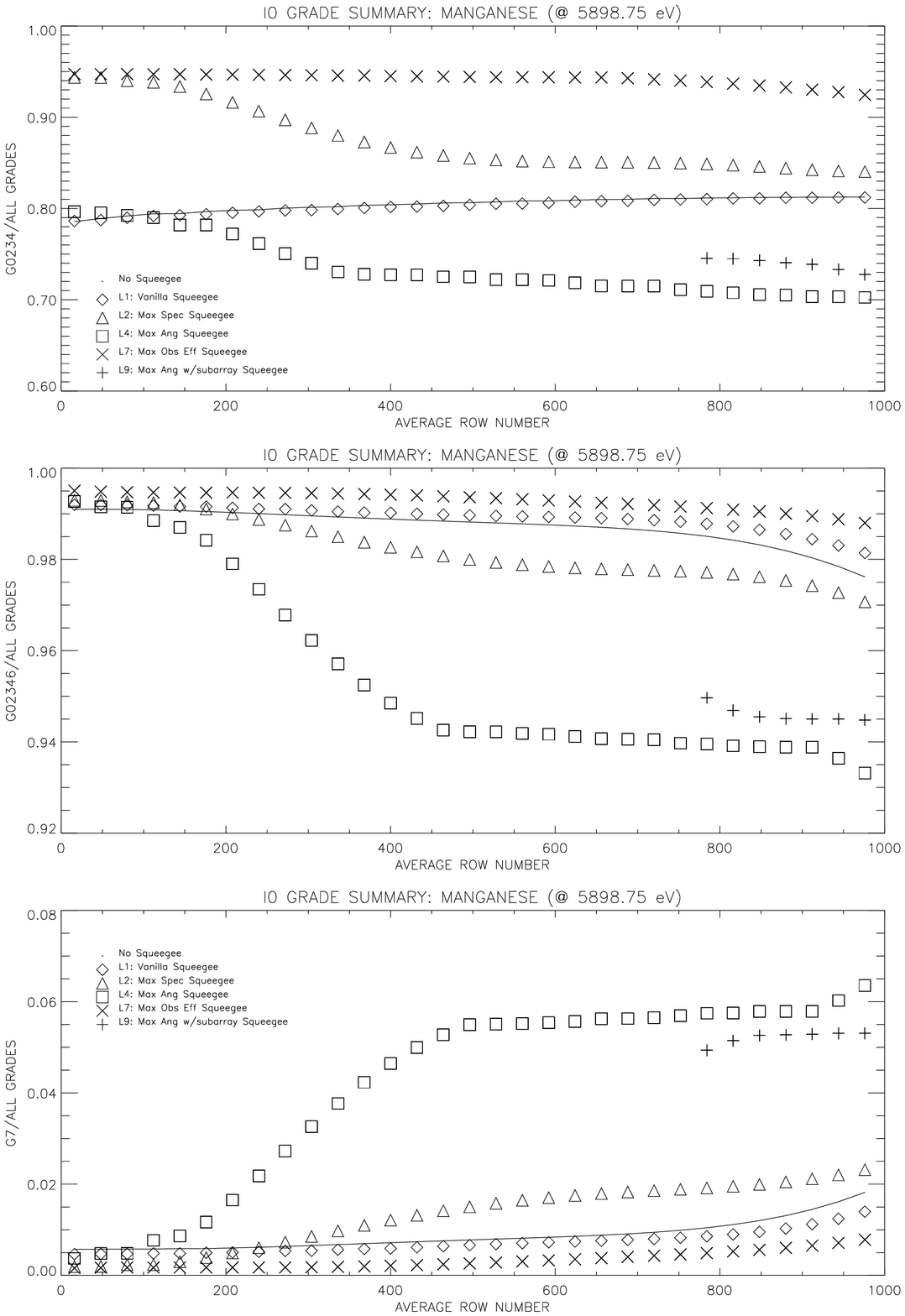}
\caption{Squeegee mode grade summary plots for the Mn-K$\alpha$ line for
   all squeegee modes.}
\end{center}
\end{figure}

The grade distribution versus number row number data are included in
Figure 4. The squeegee runs now show greater variability
with respect to each other
and a larger difference with respect to the control run.
Particularly interesting is the grade migration exhibited by squeegee
L4.  The fraction of g02346 events shows a linear decrease from
row 100 to row 400, at which point it flattens out and is much lower
than any other measurement.  This behavior is also seen in the
migration to g7 events.  
The same effect is seen in L2 but at a reduced level.
This is presumably due to the fact that L2 utilizes on-chip summing,
which should retain more of the valid X-ray events in the g02346 grade
set. The fraction of g02346 and g7 events in
the sub-array squeegee L9 lies between L2 and L4.  The shorter
frame-time of L9 compared to L4 appears to reduce some of this
migration indicating that the effect cannot be attributed solely
to the short timescale traps.

The grade migration effect seen at Al-K$\alpha$ for squeegees L2 and L4
is more pronounced at Mn-$\rm{K\alpha}$.  Analysis of the individual
grade ``branching ratios'' shows an even stronger
migration from grades g0, g3, and g4 to g2 than for Al-$\rm{K\alpha}$.
Squeegee L4 also shows
a steep increase in the g7 events from row 100 to row 500.  The flight
grade analysis shows that the migrating events are going into flight
grade 64.  It is interesting to note that the percentage of g2
events is larger in L2 than L4.  The difference between these two
squeegees is that L2 uses on-chip summing while L4 does not.
This suggests that a significant amount of charge is trailing the
center of the event by 2-3 or perhaps even more pixels. 
Lastly, L2 and L7 both have much
higher g0234$/$all grades ratios than the other modes because both L2
and L7 utilize 2x2 summing.

\subsection{Ti results}


The results for the Ti-$\rm{K\alpha}$ line are not presented in
this paper due to space restrictions. However, the Ti-$\rm{K\alpha}$ 
data for I0 resemble the Mn-$\rm{K\alpha}$ data. The grade
distribution results for Ti are also similar to the Mn results as
one would expect given that the performance characteristics of the CCD
vary little from 4.5~keV to 5.9~keV.  

\subsection{Correction for Decay of Radioactive Source}

These squeegee tests were run
between 3 and 6 months after the control run.  The half-life of the 
${\rm Fe^{55}}$ source is 2.7~yr.  Therefore, the intensity of the
source decays by $\sim6\%$ in three months and by $\sim12\%$ after
six months.   In Figure 5 we have corrected for this
decrease by simply normalizing the detected count rates of the
squeegee measurements to the control run.  After applying this
correction,
squeegees L1 and L7 have the same or higher detection efficiency than
the control run across the CCD, with the possible exception of L7 at
Al-K$\alpha$.  The upturn in the detection efficiency for the control
run for Al-K$\alpha$ is probably an artifact of the fitting process.
It is interesting that the squeegee runs show a decrease of
the detection efficiency with row number relative to the control run
at Al-K$\alpha$, but squeegees L1 and L7 match or exceed the control
run at Mn-K$\alpha$ and Ti-K$\alpha$.
L2 appears to be still lower than the control run near
the frame-store and drops more rapidly with row number than the control
run.  L4 still shows the largest drop with row number, but may now be
consistent with the control run near the frame-store.  The most
important conclusion to be drawn from these data is that squeegees
L1 and L7 {\em have the same detection efficiency as the standard
clocking} at Mn-K$\alpha$ and Ti-K$\alpha$.  

\begin{figure}[htb]
\begin{center}
\epsfxsize=3.0in
\epsfysize=3.0in
\hspace*{0pt}
\epsffile{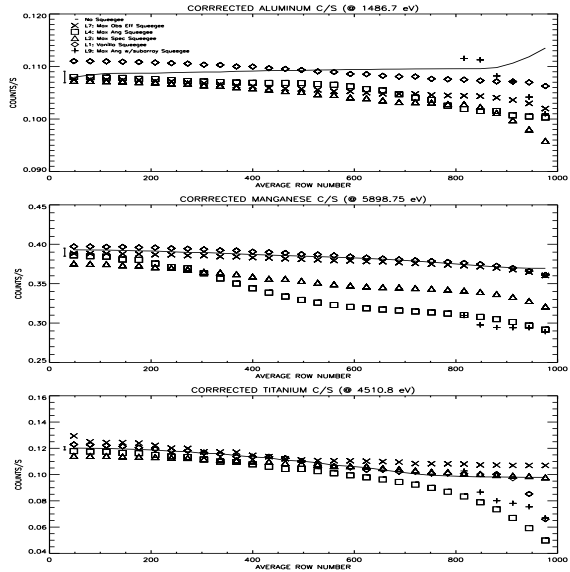}
\caption{Detection Efficiency of Squeegee Modes accounting
for radioactive decay of the calibration source.}
\end{center}
\end{figure}

  The question still remains as to the lower detection efficiency of
L2 and L4.  The detection efficiency of L4 is $\sim20\%$ lower than
the standard clocking at the top of the CCD.  Only about 6\% of this 
difference can be explained by grade migration to {\tt ASCA} g7.  
These measurements were executed with the now-standard on-board
rejection of flight grades 24, 66, 107, 214, and 255.  Of these,
flight grade 66 may be the most interesting because it is the closest
to flight grade 64 which is exhibiting the tremendous increase. 
Perhaps there is also a significant migration to grade 66, which
would lead to a lower detection efficiency since these
events are never telemetered.  It is also possible that flight 
grade 255 is enhanced by this effect.  

The variation with row number for L2 and L4 confirm what our analysis
of the S0 data indicated (Virani and Plucinsky 2000).  The detection 
efficiency is changing with
row number and the percentage decreases are the same for the 
squeegee runs which utilize the slow parallel transfer.
The severe grade migration effect appears
to lower the effective detection efficiency around row 200 for 
Mn-K$\alpha$ and row 700 for Ti-K$\alpha$.   The Al-K$\alpha$ data
are not effected until about row 900.  With the improved statistical
precision of the Ti-K$\alpha$ data, we can start to see the energy
dependence of this effect. Clearly, the Mn-K$\alpha$ photons are
effected sooner than the Ti-K$\alpha$.  This is suggestive of a
strong dependence on energy in a rather narrow range from 4.5 to
5.9~keV.

One suggested explanation for the lower detection efficiency is that
the analysis is confused by the blending of the K$\alpha$ and 
K$\beta$ lines as the spectral resolution degrades.  We note that this
effect should vary in magnitude with row number since the resolution
is degrading with row number.  However, squeegees L2 and L4 show a
discrepancy with respect to the control run in the first 200 rows.
Line blending is not an issue in the first 200 rows on I0 since the
spectral resolution is still close to pre-launch values and is more
than sufficient to resolve the K$\alpha$ and K$\beta$ lines for Mn and Ti.
Line blending may be part of the explanation near the top of the CCD
as the discrepancy between the squeegees and the control runs
increases with row number.  Nevertheless, the lower detection
efficiency near the frame-store is a puzzling effect which 
warrants more investigation. 

\section{Conclusions}

We confirm  that all of the tested squeegee modes improve
the spectral resolution of the I0 CCD compared to the standard clocking.
Our analysis of the detection uniformity indicates that 
squeegees L1 and L7 have the same detection efficiency as the
standard clocking after correcting for the decay in the intensity of
the radioactive source, while the squeegees L2, L4, and L9 still
exhibit a lower detection efficiency.  The discrepancy is as large
as 20\% for squeegee L4 at the top of the CCD. L2 and L4 also 
produce a highly spatially-dependent grade distribution.
We suggest that the slow parallel transfer of both these modes
is the likely explanation for this effect.  
We suggest that this effect should be investigated further with the 
hope that a squeegee mode can be
developed which optimizes the spectral resolution, the detection 
efficiency, and the {\em uniformity} of the detection efficiency.

\section{Acknowledgments}

	We thank our ACIS and CXC colleagues, particularly Mark Bautz,
Dan Schwartz, and Peter Ford, for many useful ideas over the course of this
analysis. SNV and PPP acknowledge support for this research from 
NASA contract NAS8-39073; CEG and BL acknowledge support for this 
research from NASA contracts NAS8-37716 and NAS8-38252.


\begin{references}

\reference Prigozhin, G., \textit{et al.}, ``Characterization of the radiation
damage in the Chandra X-ray CCDs'', SPIE Proceedings, vol. 4140,
August, 2000, pp. 123-134

\reference Bautz, M. and Kissel, S., ``Explanatory Note on Squeegee
Mode'', Internal MIT$/$ACIS Memo, 23 May 2000

\reference Bautz, M. and Grant, C., ``Choosing an ACIS Squeegee
Mode'', Internal MIT$/$ACIS Memo, 15 August 2000

\reference Virani, S. N. and Plucinsky, P. P., ``Analysis of ACIS Squeegee
Mode Data on Chip S0'', Internal CXC/ACIS Memo, 29 August 2000

\end{references}
\end{document}